\documentclass{article}

\usepackage[american]{babel}

\title{Stop the tests: Opinion bias and statistical tests}
\author{Andr\'e C. R. Martins \\
	NISC -- EACH -- Universidade de S\~ao Paulo,\\
	Rua Arlindo B\'etio, 1000, 03828--000,  S\~ao Paulo, Brazil}

\begin{document}


	\maketitle
	

	\begin{abstract}
		
		When statisticians quarrel about hypothesis testing, the debate usually focus on which method is the correct one. The fundamental question of whether we should test hypothesis at all tends to be forgotten. This lack of debate has its roots on our desire to have ideas we believe and defend. But cognitive experiments have been showing that, when we do choose ideas, we become prey to a large number of biases. Several of our biases can be grouped together in a single description, an opinion bias. This opinion bias is nothing more than our desire to believe in something and to defend it. Also, despite our feelings, believing has no solid logical or philosophical grounds. In this paper, I will show that if we combine the fact that even logic can never prove an idea right or wrong and the problems our brains cause when we pick ideas, hypothesis testing and its terminology are a recipe for disaster. Testing should have no place when we are thinking about hypothesis.

		Keywords:  hypothesis test, rationality, biases, cognition

	\end{abstract}

	%
	%
	%

	\section{Introduction}

	When Fisher \cite{fisher22a} first proposed the basic concepts of what would become hypothesis testing, the goal of finding a method to choose between ideas seemed to be a solid justification for his method. The same justification was behind Neyman's  version of testing \cite{neyman34a} . Despite both men rivalry over technical details \cite{lehmann11a}, their contribution to scientific methods was clear. Compared to much of what was done in so many fields before researchers started using statistical methods, the methodological advance was quite fundamental.
	
	However, challenges to those methods did not take long to appear. Around the same time Neyman was working on his proposal, bayesian methods\cite{definetti30a,ramsey31a} were also being defended as the correct way to conduct inference. The opposition of frequentism and bayesianism became one of the central issues in the development of statistics. But that was not a debate on whether to perform tests. Bayesian methods do not require testing. The inference can be over once you find the posterior probability for each possible theory. But Bayesian tests are possible and hypothesis tests have been proposed under the Bayesian framework \cite{pereira2011a,johnson13a}. At the same time, it is possible to do a Frequentist analysis of any problem and not use hypothesis tests. Confidence intervals and likelihood comparison do not need prior information.
	
	The debate on testing should have focused on two questions. The first one is whether it makes sense to test and possibly reject ideas at all. The second problem is, assuming that was a good idea, how it should actually be performed. Discussions of the second problem are common and they point at several technical problems with tests \cite{bayarrierger2004a,Kadane_2016a}. Yet, even when someone points out that most null hypothesis tests make no sense because the equality in null hypothesis is often known to be false, that is a technical concern. Most of the time, no comment is found on whether we should have tests. And this lack of analysis go on despite problems with the tests we have today. Unfortunately, as a community, we forgot about the first question. And we spent too much time on the second one. In this short note, I intend to address this lack of inquiry on the logical basis of hypothesis testing. More specifically, I will show that a better understanding of human cognition provides us important clues on how testing might be making us far more harm than good.

	\section{Our fallible cognition}
	
	While investigating how people composed the opinions that supported their ideas, Jervis came upon a startling observation \cite{jervis76a}. The opinions often showed too much consistency. Among the studies he compiled, there was one where the subjects about tests of new nuclear weapons. The decision whether to support those tests or not could depend on each person position on a number of underlying issues. Those positions should inform the final individual opinion on whether those tests should be done. To test how people made that type of decision, the subjects expressed their opinions on three independent positions. Maybe the tests would pose medical risks to the population, maybe not. They might cause problems by increasing international tension or not. And there was also the question on whether the tests would lead to major advances in the weapons. All those questions are relevant to the final opinion if more tests should be performed or not. But they are also independent problems. A person could hold a set of beliefs where some of the answers opposed the tests while others supported it. From there, the person would estimate which questions were more important and that should inform her final opinion. That was not what the experiments showed. Since the three issues are independent, holding one opinion on any of them should not be correlated to holding opinions on the other ones. And yet, it was observed that the opinions came in packages. People who defended the tests tended to regard them as a positive thing in every question. Those who opposed thought they would cause problems in every associated issue. It was as if people would pick the ideas that supported their final preference. Instead of reasoning about the pros and cons of the theme, their opinions were too consistent.
	
	Since we learned about  Allais \cite{allais} and Ellsberg \cite{ellsberg61} paradoxes, we have become aware that our rationality has problems. The first tests, as well as many others that followed \cite{kahnemantversky79,birnbaum2008a}, were often about non-trivial decisions under uncertainty. Since that beginning, we have been trying to understand why we make so many mistakes. A first answer to that question was proposed by Simon \cite{simon56} in the concept of satisficing. We have limited brain power. Therefore, finding answers that corresponds to the optimal solutions might not be a good strategy. It makes sense to reason in ways that need less mental effort, to save time and energy. Our mistakes wouldn't be failures, they would just represent our bounded rationality. Indeed, it is true we are limited. But we don't make mistakes only because we can not afford to spend too much time thinking about a question. If that were true our mistakes should disappear when the situation allowed us to analyze a problem with more care. And that does not seem to be the case. While part of the answer, satisficing can not account for all our mistakes. 
	
	In many situations, it seems that we use heuristics to get fast answers \cite{tversykahneman83a}. Well tuned heuristics can provide reasonable solutions and they can be quite efficient under the right circumstances \cite{gigerenzergoldstein96}. But, while they are efficient, but they can cause mistakes. And complete analysis of any problem should not rely on them when time allows and we identify something has gone wrong. Indeed, it looks like we have two reasoning systems working inside our minds \cite{kahnemanfrederick02a}. One system would be fast and take no deliberate effort to use, based on sensible but fallible heuristics. The second system would be a slower mode of thinking, requiring conscious effort. We would use it to try to correct the answers from the first system, when needed. 
	
	When facing uncertain problems, other effects also seem to be present. The way we handle uncertainty can be accounted, at least partially, by the fact that exact probabilities do not exist in the real world. Any actual probability estimate comes from observation. And, aside scientific studies, most samples in real life are often small in size\cite{kareev97}. In real life probability statements are just estimates and they should be used as data instead of exact knowledge \cite{martins05b,martins06}. By considering the uncertainty on the stated probability values, several biases can be better understood. From weighting functions \cite{kahnemantversky79,birnbaum2008a} to effects such as our failure to estimate conjunctive events \cite{cohen79}, conservatism \cite{philipsedwards66} and the effects of prior guesses on our estimate of correlations \cite{chapman69,hamiltonrose80}, it seems our brains have heuristics in place that always assume the problems are more complex than the initial experiments assumed. Our brains seem to change the probability values we hear for a simple reason. They might just be trying to correct the errors in probability estimation as well as the possibility that some deception could be involved \cite{martins06}. We might reason in ways that resemble a Bayesian analysis \cite{tenenbaumetal07}. 
	
	But accounting for external sources of error does not explain it all either. We are too confident about our own reasoning. We judge our efforts as far more likely to be correct than they are \cite{oskamp65a}. As we get more information, we become more certain even when we do not get more accurate \cite{halletal07a,tsaietal08a}. This seems to happen under most circumstances. Overconfidence was observed  even when money is involved or in professional decisions \cite{albahutchinson00a,praetoriusetal13a,chaplinshaw15a}. Interestingly, we seem to be far more overconfident about our own chances than about the same chances of others \cite{cooperetal88a}.
	
	The picture we have is one of a limited reasoner who uses fast strategies to get fast answers. Those heuristics can go wrong and often do. But they allow us to do a lot with less effort and we might even have strategies that correct for the errors of others. And some of our heuristics might exist to correct the estimates of others. In the circumstances when we do get things wrong, we might have a slower, more methodical way of thinking to correct our mistakes. But none of can explain our certainty about how good our own reasoning actually is. In this paper, I will show show that our desire and tendency to believe is the missing part of this puzzle, the main bias that encompasses many others. There is no logical support for beliefs \cite{martins16a} and we must identify our desire to defend our opinions as a bias. And, as an example, we will see that this bias is so insidious that it is the actual cause we have created statistical tests. In that sense, hypothesis testing can also be seen as part of the same bias.

	\subsection{A desire to defend ``The Truth''}
	
	Our limited abilities and the need for fast and reliable heuristics  are some of the reasons why we make so many mistakes. But if they were the sole reasons, we should not treat our own abilities as so much better than the abilities of others. 	
	Like Jervis' irrational inconsistency, many observed departures from rationality are associated  with how we handle our beliefs. We defend our points of view too strongly. And we show a clear tendency to change our opinions less than we should when we obtain new information \cite{philipsedwards66}. This effect happens even when we have just formed an initial opinion. If we get results that at first point to one option and later point to the opposite choice, the order we observed the evidence matters \cite{petersonducharme67a}. Once we have a guess about a situation, we tend to keep that guess even when the total evidence is neutral. Combine these observations with our tendency to trust ourselves far more than we should and how we assume we can have control over completely random situations \cite{langerroth75a,proninetal06a}, and a picture of a confident fool seems to emerge.
	
	If that was the total amount of evidence on how we change our minds and how overconfident we are, the confident fool would be a reasonable assumption. But, by looking at the reasoning of individuals alone, we would have missed a crucial part of our mental strategies. As social beings, human cognition problem is related to how we interact with other people. And it is related to the information others provide us and how they use the information we provide. We already know we can be heavily influenced by social pressure. Such influence can even  lead us to accept wrong answers even when they are clearly wrong \cite{asch55a,asch56a}. 
	
	But we don't treat all information from other people equally. Instead, we clearly divide things into categories. People and information that agree with us are different from those who do not. When investigating an issue, we focus our search on sources who already agree with us \cite{nickerson98a}. When presented with different pieces of information, where some agree with our beliefs and some don't, we do not treat them the same way. Data and arguments that agree with our preconceived ideas tend to be accepted at face value. When they don't agree with our believes, we actively examine the information, looking for errors and problems \cite{taberlodge06a}. And the same wrong behavior is seen under very different lights depending on who commits it. If a misdeed is performed by people from our side of a political divide, we see it as something minor. If the same action comes from people supporting views we oppose, we consider the same action a serious crime \cite{Claassen2015a}.
	
	Our preference for information from people of our group is very strong. Even 4-year old children seem to prefer an unreliable informant inside their group over a reliable outsider \cite{macdonald13a}. When reasoning about issues where we have  strong views such as  abortion \cite{luker85a} or politics \cite{searswhitney73a}, many of us seem to be incapable of even considering ideas that are opposite to our beliefs. These and similar results have led more than one researcher to challenge the concept that we reason to find the best solutions to problems. The reason why we have evolved our mental skills might be to win arguments instead of finding best answers. Our reasoning might exist to allow ourselves to convince our peers or to accept the views of our group when we can not win the argument \cite{merciersperber11a}. And this seems to be also true for children \cite{mercier11a} and different cultures \cite{mercier11b}. Our reasoning seems to exist to defend our beliefs, in particular those identity-defining beliefs we feel we must defend \cite{kahan13a}. In political situations, we can clearly see how we do not look for truth. Instead, it seems we are politically motivated \cite{kahan16a}. A particularly clear evidence that we use our brains to support our beliefs instead of verifying them comes from the fact that well educated and intelligent people seem to make simple analytical errors when the data does not agree with them  \cite{kahanetal13a}. 
	
	\subsection{Does choosing an idea make sense?}
	
	The picture that emerges is not one of incompetence. But it is one we need to worry about.  From a logical point of view, there is no known support to the concept of believing \cite{martins16a}. Deductive logic and mathematics only offer proofs when we assume initial propositions or axioms as true. They offer no actual way to choose those initial ideas. Aristotle already recognized this problem \cite{aristotles07a}. In the {\it Organon}, he proposed those initial propositions would be accepted as truth after an inductive process of reasoning. But his inductive methods, as well as every other method later proposed, were not able to offer proofs or certainty. Instead, induction is always fallible. Any ideas we can have about the real world carry a variable but not null amount of uncertainty.
	
	That does not mean that inductive methods are useless. If we abandon hope to achieve certainty, we can find ways to measure uncertainty. Induction becomes possible, but only on a probabilistic sense \cite{howsonurbach,jaynes03}. We can try to estimate how probable an idea is. But there is no known way we can make claims about the real world with certainty. On the contrary, there are many problems where induction might be so hard to do that even providing probabilities might not be possible. Uncertainty is unavoidable.

	\section{Biases in the history of statistics}
	
	The p-value crisis is not new. We know how it ties to publication bias \cite{ioannidis05a,gigerenzermarewski15a}. The widespread use of p-values has also caused severe problems due to p-hacking \cite{headetal15a}. But it is also true that p-values are not the only problem \cite{leekpeng15a}. Misuse of statistics play an important part in the current crisis. But the use of any null hypothesis significance testing procedure (NHSTP) is now considered so destructive \cite{cumming14a} that we have already seen it banned as an acceptable method \cite{trafimowmarks15a}.
	
	

	\subsection{Creating tests}
	
	The foundations of modern statistics and what we came to know as hypothesis tests were laid by Fisher in 1922 \cite{fisher22a}. In his paper, Fisher laid many of the definitions that are central to statistical thinking today. But Fisher did not focus only on hypothesis testing, even if the paper does include them as tests of significance. It is worth mentioning that he did present case against the inverse probability methods. And, that way, he set the ground for the conflict between frequentists and bayesians. From this important beginning, hypothesis testing evolved with crucial contributions from Neyman \cite{neyman34a}. While both statisticians ended up developing a strong rivalry over technical details of how testing should be done \cite{lehmann11a}, their contributions were central to the creation and advancement of the methods that we now identify as null hypothesis significance testing. 
	
	Around the same time, others were building the basic ideas of the bayesian methods\cite{definetti30a,ramsey31a}. Frequentist tests regard probability as the relative frequency a result is obtained if an experiment is identically repeated infinite times. Bayesians, on the other hand, define it as a personal measurement of the evidence for the truth of a proposition. Bayesian tests do exit \cite{johnson13a} but they are not as commonly used. This difference set the stage for the most important division in the ranks of statistics during the XX century. Frequentists refuse to accept bayesian methods because there was no way to get rid to the subjectivity in the choice of the initial opinions. Meanwhile, bayesians refuse to accept frequentist methods because they are not logically consistent. While a relevant discussion, it helped keeping the real problem with hypothesis testing hidden for too long. A question we should have asked from the beginning was if we should be performing tests at all. But we didn't know how believing and the desire to belief can compromise our mental skills \cite{martins16a}.

	\subsection{Not even wrong}
	
	Fisher and Neyman had serious criticisms about each other methods. Both of them could see how the methods of the other had subjective components \cite{lehmann11a}. They argued over the existence of subjectivity in the opponent proposal and failed to see it in their own. And both of them claimed bayesian methods were wrong due to their explicit subjectivity. All accusations of subjectivity were indeed correct. But the ones committed by the other party were always seen as a far more serious transgression. That tendency to regard the mistakes or transgressions of the opposite ideological field as more serious than our own is now a well documented effect \cite{Claassen2015a}. It certainly played an important role in the development of statistics. Each party was quite overconfident about the correction of their proposal. Testing seemed like a good idea. It has the proper appearance of a mathematical method for finding the correct answers.
	
	Yet, every single statistician will answer correctly if asked whether their methods can provide the one definitive answer.  They know (and knew) the trivial fact that none of the statistical methods can provide certainty. Whatever your favorite choice, the best you can hope to say is that the evidence seems to favor one hypothesis (if you are frequentist). Or that one hypothesis is more probable than the other one (if you are a bayesian). When a statistician uses a test, she knows that the answer it provides might be wrong. This is very clear, the types of errors should make that obvious to any student. But the terminology still mentions rejecting hypothesis. Even while it is clear that the mentioned hypothesis might be true.
	
	It seems that our human biases played a major role in the creation of the testing framework. Our desire for control, our tendency to look for something to believe and defend are always with us. That means that tools that can provide us answers we can defend are tools we would welcome. Statistical tests, regardless of whether they are done in a frequentist or bayesian way, provide those tools. They are our best answer to our desire to look scientific while accepting and rejecting ideas. They allow us to reject ideas despite the fact it is clear that we do not have the answers, that randomness is present. Tests admit that as chances for errors. This admission has protected them from any claim that they are wrong. Indeed, they are not wrong. After all, the whole framework admits there are chances that mistakes will be made. And that is a crucial point. But their results make us easy prey to our desire to know, even when knowledge is not warranted. We trade being correct for being sure.
	
	Testing seems to be a product of our opinion bias. When they were created, we didn't know about the deleterious effects associated to how we reason and how we defend our opinions. The task of looking for methods to establish beliefs that looked solid from a mathematical point of view seemed to make sense. Indeed, the same error can also be found in another normative field for the scientific endeavor, philosophy of science. Take Popper, for example. While admitting we can not know an idea is true, he assumed they could be rejected \cite{popper59}. This is the exact same framework. You can not know your null hypothesis is true but you can try to reject it. In the philosophy of science, we have already learned that rejecting theories is also not feasible. Any prediction depends on auxiliary hypothesis \cite{duhem89a,quine69}. When an experiment seems to suggest a theory should be discarded, it is possible that the problem is due to the auxiliary hypothesis and not the theory. Theories can not really be falsified from a logical, non-inductive point of view. But our desire to defend ideas is so strong that even in the bayesian literature we will find schemes to decide which theory to choose \cite{maher93}.
	
	Tests provide you an answer when you choose an acceptable chance of being wrong. But the scientific endeavor is not about being wrong sometimes. While that is unavoidable, our tools should be about finding those times when we are wrong, about refining our knowledge. Of course, decisions are often necessary. But decisions are about action, not about choosing an idea as THE truth when there is no logical proof. Real scientific work includes observing the world and proposing ideas to explain it. Scientific work ought to be about checking all possible explanations and creating new ones nobody has thought before \cite{martins16a}. And using all the data we cano see which ones are more likely true, not which ones are indeed true. 
	
	The picture that emerges is clear. Statisticians, as everyone else, are subject to our human biases. Our opinion bias has influenced them to create the tools we have been using. As honest and intelligent researchers, the statistical pioneers realized from the beginning that no final answer could be achieved. But the notion that we, as humans, can actually learn the one true answer was too deep inside their minds. They wanted to support their opinions. Even when those opinions could just be wrong. That desire is at the root of the devastating effects in scientific reliability we are now observing. Publication biases, p-hacking, and the idea that our statistical tools could provide truth are linked problems. There are situations where probabilities are so close to zero that we may dismiss some ideas as irrelevant. If you compare how the predictions of Einstein and Newton's gravity with data, you do find that the probability that Newton's ideas are correct have so many zeros that you would have to be a mathematician to call it anything other than zero \cite{martins16a}. In those cases, it makes sense to abuse the language and say a theory has been disproved. In every day experiments, choosing a side makes no real sense.
	
	It is interesting to notice that, used with the proper caution, we can use a test as a preliminary measurement to decide whether to pursue a certain investigative path or not. Those decisions are often based on guesses and a test can provide better information than a simple guess. Tests should have the same status as our heuristics have in human reasoning. Useful and fast tools that we use for fast decisions but, when serious analysis is required, we must pause and use a more complete and correct approach.

	\section{Discussion}
	
	Hypothesis testing was created to allow us to verify which ideas seem to be true. But in the problems where it can be applied, no claim of truth is possible. Inductive methods can at best provide us probability estimates. There are situations when the data is such that a hypothesis is extremely unlikely to be true. But almost certain is qualitatively different from real certainty. That means that, at the very least, we should change the terms we use. Rejection and acceptance suggest a knowledge that is a lie. A test does not reject a hypothesis. It shows, at best, that the hypothesis is unlikely.
	
	There are practical problems that look, at a first glance, as if they require a choice between ideas. But that is actually never true. When a physician faces a difficult diagnosis, it might seem as if she has to choose between possible theories that match the patient symptoms. But what she actually has to choose is her action, and the action is which treatment she will suggest. That depends on which condition is more likely and the consequences of making mistakes. But she should not decide about the disease. She can consider one as far more likely, but she should keep her mind open to the possibility that new data will change that. We choose actions, we should never choose ideas about the real world. Your preferences are yours but they are about what you would like and not about how the world actually is. Descriptions of the real world, on the other hand, are subject to uncertainty. And we need our methods to keep that clear. Limited cognition, conflicting paths, there are instances when we will be forced to choose an action. That choice might naively look as accepting or rejecting ideas. But they are actually choosing actions, subject to decision theory arguments. When a teacher chooses the best description to teach her students, the limiting factor is time. Otherwise, all competing theories should be presented, the more probable ones mentioned as such. When we launch a rocket to the Moon, we might even decide to use Newtonian Mechanics over General Relativity despite the data, because calculations will provide the same basic results and it is easier to perform them using Newtonian Mechanics. It seems we are choosing ideas in both case, but, far from it, we are using normal decision arguments to choose actions. But the notion we can rejecting hypothesis that might still be true should be completely abandoned.
	
	Statistics needs to become conscious of our reasoning limitations. It exists and is amazingly useful because our brains can not reach reliable conclusions from data. Statistical methods, therefore, should be planned to avoid the traps of our reasoning. When they make those traps more dangerous, it is time to change methods. Above all, statistical tools exist to assist and correct our limited and fallible reasoning. Not to reinforce the biases we already have.

	\section*{Acknowledgments}
	The author would like to thank the Funda\c{c}\~ao de Amparo \`a Pesquisa do Estado de S\~ao Paulo (FAPESP) for partial support to this research under grant 2014/00551-0.
	
	\bibliography{biblio}    
	\bibliographystyle{unsrt}

\end{document}